\pdfoutput=1
\documentclass[aps,pre,twocolumn, superscriptaddress]{revtex4-2}

\usepackage{graphicx}
\usepackage{amsmath}
\usepackage{amssymb}
\newcommand{\RN}[1]{\textup{\uppercase\expandafter{\romannumeral#1}}}
\usepackage[subrefformat=parens,labelformat=parens,caption=false]{subfig}
\begin{document}

\preprint{1}
\title{Algebraic equations of state for the liquid crystalline \\ phase behavior of hard rods}





\author{V.~F.~D. Peters}
\affiliation{Laboratory of Physical Chemistry, Department of Chemical Engineering and Chemistry \& Institute for Complex Molecular Systems, Eindhoven University of Technology, P.O. Box 513, 5600 MB Eindhoven, The Netherlands}

\author{M. Vis}
\affiliation{Laboratory of Physical Chemistry, Department of Chemical Engineering and Chemistry \& Institute for Complex Molecular Systems, Eindhoven University of Technology, P.O. Box 513, 5600 MB Eindhoven, The Netherlands}
\affiliation{%
Laboratoire de Chimie, \'Ecole Normale Sup\'erieure de Lyon, 69364 Lyon CEDEX 07, France
}%

\author{H.~H. Wensink}
\affiliation{Laboratoire de Physique des Solides - UMR 8502, CNRS \& Universit\'{e} Paris-Saclay, Orsay, France}

\author{R. Tuinier}
\email[]{r.tuinier@tue.nl}
\affiliation{Laboratory of Physical Chemistry, Department of Chemical Engineering and Chemistry \& Institute for Complex Molecular Systems, Eindhoven University of Technology, P.O. Box 513, 5600 MB Eindhoven, The Netherlands}
\affiliation{Van ’t Hoff Laboratory for Physical and Colloid Chemistry, Department of Chemistry \& Debye Institute for Nanomaterials Science, Utrecht University, 3584
CH Utrecht, The Netherlands}

\date{\today}

\begin{abstract}
Based on simplifications of previous numerical calculations [Graf and L\"{o}wen, Phys. Rev. E \textbf{59}, 1932 (1999)], we propose algebraic free energy expressions for the smectic-A liquid crystal phase and the crystal phases of hard spherocylinders.
Quantitative agreement with simulations is found for the resulting equations of state.
The free energy expressions can be used to straightforwardly compute the full phase behavior for all aspect ratios and to provide a suitable benchmark for exploring how attractive interrod interactions mediate the phase stability through perturbation approaches such as free-volume or van der Waals theory.
\end{abstract}


\maketitle



\section{Introduction}
Viruses often have rod-like shapes and can display a variety of lyotropic liquid crystal phases, as found from studies on dispersions of tobacco mosaic virus \cite{Dogic1997} or the bacteriophage feline distemper \cite{Grelet2014}.
Similar liquid crystal phases have been studied in synthetic systems of rod-like boehmite or silica colloidal dispersions \cite{Buining1993,Kuijk2012}.
To understand the role of particle shape and configurational entropy on the stability of these colloidal phases, it is useful to examine a system of hard-core particles, where volume exclusion between the cores prohibits particle overlap without the presence of additional soft interactions.

\begin{figure}
\includegraphics[width=\linewidth]{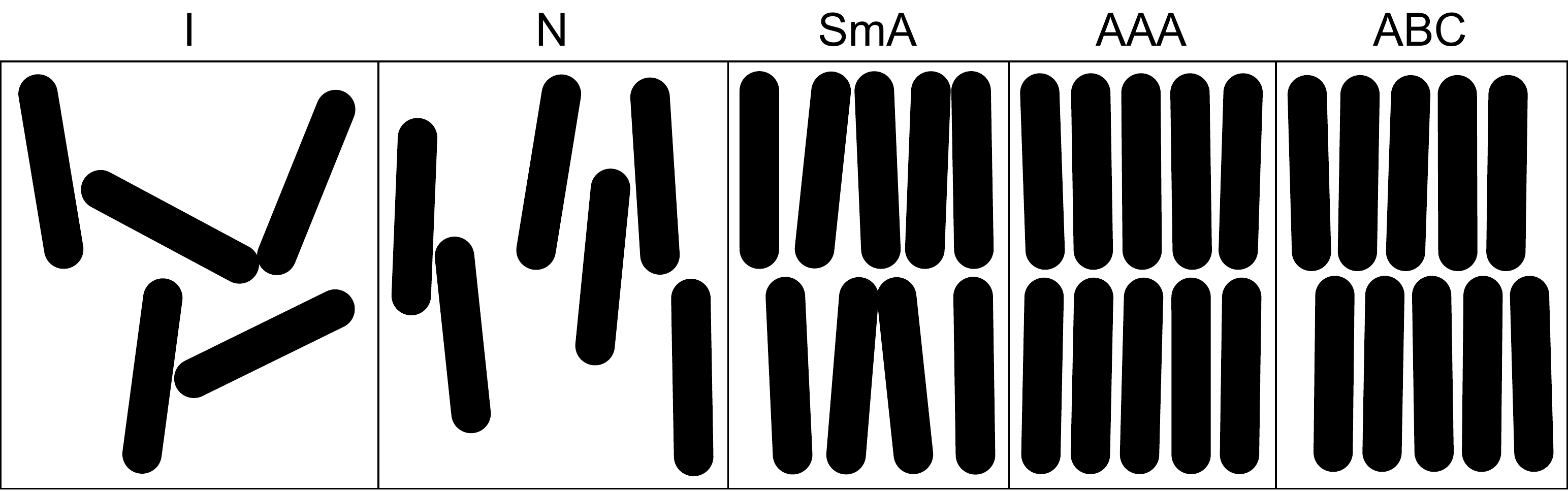}
\caption{Schematic images of the isotropic (I), nematic (N), smectic-A (SmA), AAA crystal, and ABC crystal phase states of hard spherocylinder suspensions.}
 \label{fig:gridphase}
\end{figure}

For monodisperse hard spherocylinders Monte Carlo simulations have revealed the emergence of isotropic, nematic, smectic-A and crystal phases as the concentration is increased (see Figure \ref{fig:gridphase}) \cite{Frenkel1988,Mcgrother1996,Bolhuis1997}.
In the isotropic and nematic phases the particles can freely move in all directions, while there is a preferred orientation of the particles  in the nematic phase \cite{DeGennes1974}.
The smectic-A phase consists of particles that are roughly confined in layers wherein the rods are aligned normal to the layer and diffuse laterally thus displaying the behavior of  a crowded liquid.  
The crystal phases are characterized by a similar lamellar organization but with the rods exhibiting long-ranged hexagonal order across the layer.
In the AAA crystal phase rods are stacked directly on top of each other, while for the ABC crystal they are stacked in between the rods of the adjacent layers.
The stacking in the ABC crystal is therefore equivalent to that of an FCC crystal.
For relatively short rods the nematic, the smectic-A, and AAA phases become metastable.
Other liquid crystal phases such as the smectic-B and columnar phases have been reported experimentally and their stability is attributed to additional interactions, polydispersity or semiflexibility  \cite{Grelet2014,Kuijk2012,Wensink2007,Grelet2008,Grelet2016,Paineau2016,DeBraaf2017}.

While significant progress has been made on developing predictive theories for the isotropic and nematic fluid stability, an accurate thermodynamic description of the SmA and crystal phases remains a challenging problem.  \cite{Mederos2014}.
Density functional theory \cite{Poniewierski1990,Somoza1990,Velasco2000,Wittmann2014,Wittmann2016} has proven a powerful but technically involved theoretical framework providing good agreement with simulation results for the isotropic--nematic and nematic--smectic-A phase transitions.
As a much simpler alternative to density functional theory, extended cell theory provides reasonable agreement for the full phase behavior of short rods involving crystal phases \cite{Graf1999}.
Nonetheless, both theoretical methods rely heavily on non-analytical expressions of the excess free energy, which have to be evaluated numerically.

For the columnar liquid crystal phase of colloidal platelets an analytical scaling expression for the free energy was obtained from an extended cell theory \cite{Wensink2004,Wensink2009} and the predicted phase behavior was found to agree well with computer simulation results.
This expression also enabled the use of free volume theory to determine the phase behavior of mixtures of suspensions containing plates and non-adsorbing polymers \cite{GonzalezGarcia2018}.
Inspired by this approach, we aim to seek analytical free energy expressions for the smectic-A and crystal phases of rods from an extended cell theory and map out the complete phase diagram of rod suspensions.
This approach, being entirely algebraic, considerably reduces the complexity and computational cost involved in determining the smectic-A, AAA, and ABC equations of state and associated crystal spacing.
Our algebraic theory provides a suitable starting point towards more extended approaches based on perturbation or free volume theories, aimed at incorporating soft rod interactions \cite{Franco-Melgar2008,Franco-Melgar2009}, depletion effects \cite{Lekkerkerker2011} and particle semiflexibility \cite{VanWesten2015}.

\section{Theory}
\subsection{Formulation of the free energy of hard spherocylinders}
Onsager's treatment of the entropy of anisotropic (hard) particle dispersions is the foundation of numerous theories for liquid crystal phases \cite{Onsager1949}.
Based on his definition, the Helmholtz free energy $F$ of a system of hard spherocylinders with length $L$ and diameter $D$ can be written in terms of the following entropic contributions \cite{Onsager1949}:
\begin{equation}
f=\frac{Fv_0}{Vk_\mathrm{B}T}=f_\mathrm{id}+f_\mathrm{or}+f_\mathrm{pack}.
\label{eq:ftot}
\end{equation}
Here $v_0=\pi D^3/6+\pi D^2L/4$ is the spherocylinder volume, $V$ the system volume, $k_\mathrm{B}$ Boltzmann's constant, and $T$ temperature.

The ideal free energy $f_\mathrm{id}$ is given by $f_\mathrm{id}=\eta\ln{(\eta\Lambda^3/v_0)}-\eta$, with $\Lambda$ the de Broglie wavelength and $\eta=v_0\rho$ the rod volume fraction with $\rho$ the number density of rods.
The orientational free energy $f_\mathrm{or}$ is determined by the orientational entropy, while the packing free energy $f_\mathrm{pack}$ depends on the translational entropy the rods experience.
Both depend on the probability of the particle to adopt a certain orientation, described by the orientational distribution function $\psi(\bold{\Omega})$ with $\bold{\Omega}$ the solid angle.
The function $\psi(\bold{\Omega})$ is normalized as follows:
\begin{equation}
\int \psi(\bold{\Omega})\mathrm{d}\bold{\Omega}=1.
\label{eq:norm}
\end{equation}
Since for an isotropic phase all orientations are equally probable, the orientational distribution function is a constant: $\psi=1/(4\pi)$.
For ordered phases, the rods have a preferred direction and $\psi(\bold{\Omega})$ can be found by a functional minimization of the total free energy with respect to $\psi(\bold{\Omega})$ or algebraically through the use of a trial function that depends on a single variational parameter.
The orientational free energy $f_\mathrm{or}$ per particle is related to $\psi(\bold{\Omega})$ by the following expression \cite{Onsager1949}:
\begin{equation}
\frac{f_\mathrm{or}}{\eta}=\int \psi(\bold{\Omega})\ln{[4\pi \psi(\bold{\Omega})]}\mathrm{d}\bold{\Omega},
\label{eq:frot}
\end{equation}
which for an isotropic phase leads to $f_\mathrm{or}=0$.
The approach to obtain the packing free energy $f_\mathrm{pack}$ depends on the phase state, as detailed in the following sections.

\subsection{Isotropic and nematic phase}
The free energy of the fluid phases without long-ranged positional order, the isotropic and nematic phases, was described by Onsager up to the second virial term, which is proportional to the orientationally averaged excluded volume \cite{Onsager1949}.
This gives an \textit{exact} solution for rod dispersions in the limit of infinitely long and thin rods ($L/D\to\infty$).
At finite $L/D$ rods are commonly represented as spherocylinders, i.e. cylinders equipped with a hemispherical endcap at either tip, for which higher-order virial terms need to be somehow included. 
This can be done using the approximate Scaled Particle Theory (SPT) or Parsons--Lee (PL) equations of state, which provide reasonably accurate approximations of $f_\mathrm{pack}$ \cite{Cotter1977,Parsons1979,Lee1987,Lee1988}.
The SPT and PL expressions of $f_\mathrm{pack}$ are respectively \cite{Cotter1977,Parsons1979,Lee1987,Lee1988}:
\begin{equation}
\frac{f_\mathrm{pack,SPT}}{\eta}=-\ln{\left(1-\eta\right)} + a\frac{\eta}{1-\eta}+\frac{1}{2}b\frac{\eta^2}{(1-\eta)^2},
\label{eq:nSPT}
\end{equation}
and
\begin{equation}
\frac{f_\mathrm{pack,PL}}{\eta}=\frac{4\eta-3\eta^2}{4(1-\eta)^2}\left(4+z\frac{3(\Gamma-1)^2}{3\Gamma-1}\right),
\label{eq:nPL}
\end{equation}
with
\begin{align*}
a&=3+z\frac{3(\Gamma-1)^2}{3\Gamma-1},\\
b&=\frac{12\Gamma(2\Gamma-1)}{(3\Gamma-1)^2}+z\frac{12\Gamma(\Gamma-1)^2}{(3\Gamma-1)^2},\\
z&=\frac{4}{\pi}\int\int \psi(\bold{\Omega})\psi(\bold{\Omega'})|\sin{\gamma}|\mathrm{d}\bold{\Omega}\mathrm{d}\bold{\Omega'}.
\end{align*}
Here $\Gamma=L/D+1$ and $\gamma$ is the angle between two spherocylinders with solid angles $\Omega$ and $\Omega'$.

For an isotropic phase $z$ reduces to $1$, while for the nematic phase it is more complex and an expression for $\psi(\bold{\Omega})$ is required.
Minimization of the free energy can be done numerically \cite{VanRoij2005,Tuinier2007} or using a trial function which enables an analytical solution.
While an accurate trial function has been proposed by Onsager in his original paper \cite{Onsager1949, Franco-Melgar2008}, we will use the simpler Gaussian distribution introduced by Odijk following from the limit of strongly aligned rods (large $\kappa$) \cite{Odijk1985,Odijk1986}:
\begin{align}
\psi(\theta)\approx
\begin{cases}
(\kappa/4\pi)\exp{[-\left(1/2\right) \kappa \theta^2]}&0\leq\theta\leq\pi/2,\\
(\kappa/4\pi)\exp{[-\left(1/2\right) \kappa (\pi-\theta)^2]}&\pi/2\leq\theta\leq\pi,
\end{cases}
\label{eq:nODF}
\end{align}
where $\theta$ is the polar angle with respect to the director of the nematic phase and the parameter $\kappa$, related to the width of the distribution, is found from the leading order asymptotic expressions for the orientational averages in $f_\mathrm{or}$ and $f_\mathrm{pack}$ and subsequent minimization $\partial f/\partial\kappa=0$.
This leads to the following expressions:
\begin{align}
\frac{f_\mathrm{or}}{\eta}&\approx\ln{\kappa}-1,\\
z&\approx\frac{4}{\sqrt{\pi \kappa}},\\
\kappa_\mathrm{SPT}&\approx\frac{9(\Gamma-1)^4}{4\pi(3\Gamma-1)^2}\left(\frac{4\eta}{1-\eta}+\frac{8\Gamma}{3\Gamma-1}\frac{\eta^2}{(1-\eta)^2}\right)^2,\\
\kappa_\mathrm{PL}&\approx\frac{9(\Gamma-1)^4}{4\pi(3\Gamma-1)^2}\frac{(4\eta-3\eta^2)^2}{(1-\eta)^4}.
\label{eq:nfrot}
\end{align}

To partially correct for the approximate nature of the Gaussian $\psi(\theta)$ a value of $-0.139$ is added to $f/\eta$ for the nematic phase to improve the comparison with simulation results \cite{Tuinier2016}.
This value is the free energy difference in the Onsager limit ($L/D\to\infty$) between the Gaussian approximation and the exact numerical result \cite{Lekkerkerker1984}.
The total free energy expression of the isotropic and nematic phase thus becomes \cite{Tuinier2016}:
\begin{align}
f_\mathrm{I}&=f_\mathrm{id}+f_\mathrm{pack}
\label{eq:iftot},\\
f_\mathrm{N}&=f_\mathrm{id}+f_\mathrm{or}+f_\mathrm{pack}-0.139\eta
\label{eq:nftot}.
\end{align}

Upon comparing the resulting osmotic pressure and phase behavior with simulation results \cite{Mcgrother1996,Bolhuis1997}, we found that the SPT approximation is the most accurate for long rods (Onsager limit), while the PL approximation is the most accurate for short rods (sphere limit).
Therefore we have used the following interpolation ansatz:
\begin{equation}
f_\mathrm{I/N}=g f_\mathrm{I/N,SPT}+(1-g)f_\mathrm{I/N,PL},
\label{eq:intftot}
\end{equation}
with the sigmoidal function:
\begin{equation}
g=\frac{1}{1+e^{\Gamma_\mathrm{t}-\Gamma }},
\label{eq:intg}
\end{equation}
where $\Gamma_\mathrm{t}=6$ represents the typical transition value connecting the Onsager limit ($\Gamma\to\infty$) and the sphere limit ($\Gamma\to1$).

\subsection{Smectic-A Phase}
Graf and L\"{o}wen \cite{Graf1999} numerically solved an extended cell theory model for the smectic-A phase.
In their model, spherocylinders are assumed to be confined in discrete layers with spacing $\Delta_\bot$ while the particles can freely move within these layers.
The free energy of the smectic-A phase is split into the following terms:
\begin{align}
f_\mathrm{SmA}&=f_\mathrm{id}+f_\mathrm{or}+f_\mathrm{pack},\label{eq:FtotalSPT}\\
&=f_\mathrm{or}+f_\parallel+f_\bot,
\label{eq:Ftotalalt}
\end{align}
where $f_\parallel$ is the free energy related to the fluid-like behavior in the two dimensions parallel to the layers and $f_\bot$ is the free energy related to the positional order in the dimension orthogonal to the layers.
We first consider $f_\parallel$ and $f_\bot$ separately after which we minimize the total free energy with respect to $\psi(\bold{\Omega})$ and the layer spacing $\Delta_\bot$.

In the case of perfectly aligned rods, the equation of state of a 2D fluid of hard discs with diameter $D$ describes the pressure $\Pi_\mathrm{2D}$ in the dimensions parallel to the layers.
While there is no exact expression for this fluid phase, there are accurate approximations for the entire concentration range \cite{Mulero2009}.
The following simple scaled particle theory result is used \cite{Helfand1961}:
\begin{equation}
\frac{\Pi_\mathrm{2D}}{\rho_\mathrm{2D} k_\mathrm{B}T}=\frac{1}{\left(1-\eta_\mathrm{2D}\right)^2}.
\label{eq:EOS2D}
\end{equation}
where $\eta_\mathrm{2D}=a_0 \rho_\mathrm{2D}$ is the area fraction of spherocylinders with $\rho_\mathrm{2D}$ the number of particles per unit area in a smectic layer and $a_0=\pi D^2/4$ the particle area.
This expression is accurate up to $\eta_\mathrm{2D}\approx0.7$, close to the fluid--solid transition of hard discs, where the compressibility from SPT only deviates less than $3 \%$ from simulation results \cite{Kolafa2006}.
The area fraction is related to the volume fraction $\eta$ in the following way:
\begin{equation}
\eta=\frac{v_0}{a_0}\frac{\eta_\mathrm{2D}}{\Delta_\bot}=\frac{3\Gamma-1}{3\bar{\Delta}_\bot}\eta_\mathrm{2D},
\label{eq:n2D}
\end{equation}
where $\bar{\Delta}_\bot=\Delta_\bot/D$ and $\Gamma=L/D+1$.
This leads to the following free energy $f_\parallel$ for perfectly aligned spherocylinders:
\begin{equation}
\frac{f_{\parallel\mathrm{,al}}}{\eta}=\ln{\left(\frac{\eta_\mathrm{2D}\Lambda^2}{a_0}\right)}-1-\ln{\left(1-\eta_\mathrm{2D}\right) +\frac{\eta_\mathrm{2D}}{1-\eta_\mathrm{2D}}}.
\label{eq:F2DSPT}
\end{equation}

Upon accounting for rotations of the rods, the excluded area projected by a single rod should increase.
We define $a_\mathrm{eff}$ as the effective projected lateral area occupied by each spherocylinder within the layer.
The size $D_\mathrm{eff}$ could be interpreted as the orientationally averaged minimal (i.e. at particle contact) center-of-mass distance between spherocylinders.
Previously, a definition based on the orientationally averaged width of a spherocylinder in the plane of the layers was used \cite{Graf1999}:
\begin{equation}
\frac{D_\mathrm{eff}}{D}=\bar{D}_\mathrm{eff}=1+(\Gamma-1)\int \psi(\bold{\Omega})|\bold{\Omega}\cdot\bold{\Omega}_{\theta=\pi/2}|\mathrm{d}\bold{\Omega}.
\label{eq:Dint}
\end{equation}
Here $|\bold{\Omega}\cdot\bold{\Omega}_{\theta=\pi/2}|$ is the dot product of the solid angles $\bold{\Omega}$ and $\bold{\Omega}_{\theta=\pi/2}$, where $\theta$ is the polar angle with respect to the director of the smectic-A phase.
However, this definition does not take into account configurations of other spherocylinders and thus overestimates $\bar{D}_\mathrm{eff}$.
The resulting overestimated loss of entropy and other discrepancies were corrected by adding a (negative) free energy term based on comparisons with simulation results \cite{Bolhuis1997}.

A similar problem appears for the columnar phase of plates upon quantifying an effective length $\bar{L}_\mathrm{eff}$ of the plates confined in hexagonal tubes \cite{Wensink2004,Wensink2009}.
Likewise a single particle integral was used, but to compensate for the other particles and the averaging over the azimuthal angle a prefactor `1/2' was introduced in front of the integral.
The benefit of adding a correction in the definition of $\bar{L}_\mathrm{eff}$ or $\bar{D}_\mathrm{eff}$ opposed to the addition of a free energy term is that it leads to more realistic values for $\psi(\bold{\Omega})$ and $f_\mathrm{or}$ that are comparable to those for the nematic phase.

Thus, we choose to rescale the integral from Eq.\ \ref{eq:Dint} by a factor $A$ leading to the following definition of the effective rod diameter $\bar{D}_\mathrm{eff}$:
\begin{equation}
\bar{D}_\mathrm{eff}\approx1+ A(\Gamma-1)\int \psi(\bold{\Omega})|\sin(\theta)|\mathrm{d}\bold{\Omega},
\label{eq:Deff}
\end{equation}
where $A$ was chosen such as to fit the resulting equations of state and nematic--smectic-A phase transitions to those obtained from computer simulations \cite{Mcgrother1996,Bolhuis1997}.
This means the factor $A$ varies depending on whether the equations of state for the nematic phase is based on SPT ($A=0.41\eta$) or PL ($A=0.28\eta$) and hence we have used the interpolation $A=0.41\eta h$ with $h=g+(1-g)0.28/0.41$.
As the smectic-A phase is expected to be the preferred phase state near $\eta\sim0.4-0.6$, $A$ attains values near $0.1-0.2$, which is significantly smaller than the factor $A=1/2$ proposed for plates.
The difference may be related to the additional degree of freedom within the confined layers as opposed to the confined hexagonal tubes leading to a relatively larger number of configurations of the particles at shorter distances.
Unrelated discrepancies in the free energy from for instance the penetration of rods in other layers would also influence the value of $A$.
Additionally, it was found that instead of taking $A$ as a constant the inclusion of linear $\eta$-dependence led to an improvement especially when comparing the resulting equations of state with computer simulation results.
The free energy of a system of effective 2D discs can be obtained from Eq.\ \ref{eq:F2DSPT} by substituting $a_0$ with $a_\mathrm{eff}$ and $\eta_\mathrm{2D}$ with $\eta_\mathrm{2D}a_\mathrm{eff}/a_0=\eta_\mathrm{2D}\bar{D}_\mathrm{eff}^2$ \cite{Graf1999}:
\begin{equation}
\begin{split}
\frac{f_\parallel}{\eta}&=\ln{\left(\frac{\eta_\mathrm{2D}\Lambda^2}{a_0}\right)}-1\\&\qquad-\ln{\left(1-\eta_\mathrm{2D}\bar{D}_\mathrm{eff}^2\right)}+\frac{\eta_\mathrm{2D}\bar{D}_\mathrm{eff}^2}{1-\eta_\mathrm{2D}\bar{D}_\mathrm{eff}^2}.
\end{split}
\end{equation}

For the dimension orthogonal to the layers, we consider a 1D lattice with a lattice spacing $\Delta_\bot$.
From cell theory \cite{Lennard-Jones1937} it follows that the free energy is related to the free space available to the rod.
As the rod is confined in a cell of length $\Delta_\bot$, the free space is simply given by $\Delta_\bot-L-D$.
The free energy in the dimension orthogonal to the layers thus becomes \cite{Graf1999}:
\begin{equation}
\frac{f_\bot}{\eta}=\ln\frac{\Lambda}{D}-\ln{\left(\bar{\Delta}_\bot-\Gamma \right)}.
\label{eq:FCT}
\end{equation}

The total free energy expression can now be obtained from Eq.\ \ref{eq:Ftotalalt}.
Recalling the different entropic contributions in Eq.\ \ref{eq:FtotalSPT} we write the packing free energy as follows:
\begin{equation}
\begin{split}
\frac{f_\mathrm{pack}}{\eta}&=-\ln{\left(1-\eta_\mathrm{2D}\bar{D}_\mathrm{eff}^2\right)} \\
&\qquad+\frac{\eta_\mathrm{2D}\bar{D}_\mathrm{eff}^2}{1-\eta_\mathrm{2D}\bar{D}_\mathrm{eff}^2}-\ln{\left(1-\Gamma/\bar{\Delta}_\bot \right)}.
\end{split}
\label{eq:FSPT}
\end{equation}
Now $f_\mathrm{or}$, $\bar{D}_\mathrm{eff}$, and $\bar{\Delta}_\bot$ can be determined by simultaneously minimizing the total free energy with respect to $\psi(\bold{\Omega})$  and $\bar{\Delta}_\bot$.

\subsubsection*{Free energy minimization}
 \label{sec:minF}
Given that the free energy only depends on single particle orientational integrations, it is possible to carry out the minimization with respect to $\psi(\bold{\Omega})$ analytically.
The minimization equation reads as follows:
\begin{equation}
\begin{split}&
A(\Gamma-1)
\left[\frac{4\eta_\mathrm{2D}\bar{D}_\mathrm{eff}}{\left(1-\eta_\mathrm{2D}\bar{D}_\mathrm{eff}^2\right)} 
+\frac{2\eta_\mathrm{2D}^2\bar{D}_\mathrm{eff}^3}{(1-\eta_\mathrm{2D}\bar{D}_\mathrm{eff}^2)^2}\right]|\sin{\theta}|\\
&\qquad+\log{\left[4\pi \psi(\theta)\right]}
-\lambda=0,
\end{split}
\end{equation}
where $\lambda$ is the Lagrange multiplier ensuring the normalization of $\psi(\theta)$ (\textit{cf.} Eq. \ref{eq:norm}).
This leads to the following expression for the orientational distribution function $\psi(\theta)$:
\begin{align}
\psi(\theta)=Z^{-1}\exp{\left[-\kappa|\sin{\theta}|\right]},
\label{eq:ODF}
\end{align}
with
\begin{equation}
\kappa=A(\Gamma-1)
\left[\frac{4\eta_\mathrm{2D}\bar{D}_\mathrm{eff}}{\left(1-\eta_\mathrm{2D}\bar{D}_\mathrm{eff}^2\right)} 
+\frac{2\eta_\mathrm{2D}^2\bar{D}_\mathrm{eff}^3}{(1-\eta_\mathrm{2D}\bar{D}_\mathrm{eff}^2)^2}\right],
\label{eq:kappa}
\end{equation}
and
\begin{align}
Z=\int{\exp{\left[-\kappa|\sin{\theta}|\right]}\mathrm{d}\bold{\Omega}}.
\end{align}

Similar to the nematic phase, the spherocylinders are strongly aligned so that $\kappa\gg 1$.
Therefore, we again retain only the leading order contribution for $\kappa\gg1$, which is an exponential distribution:
\begin{align}
\psi(\theta)\approx
\begin{cases}
\left(\kappa^2/4\pi\right)\exp{(-\kappa \theta)}&0\leq\theta\leq\pi/2,\\
\left(\kappa^2/4\pi\right)\exp{(- \kappa (\pi-\theta))}&\pi/2\leq\theta\leq\pi.
\end{cases}
\label{eq:ODFsim}
\end{align}
This leads to
\begin{align}
\bar{D}_\mathrm{eff}&\approx1+A(\Gamma-1)\frac{2}{\kappa}=1+\xi ,
\label{eq:Deffsim}\\
\frac{f_\mathrm{or}}{\eta}&\approx  2\ln{\kappa}-2,
\label{eq:frotsim}
\end{align}
where $\xi$ can be interpreted as the effective increase in diameter.
It should be noted that for a columnar phase of platelets with an effective length $\bar{L}_\mathrm{eff}$ the same form of Eqs.\ \ref{eq:ODFsim}--\ref{eq:frotsim} was obtained \cite{Wensink2004,Wensink2009}.

The value of $\xi$ can be found by inserting Eq.\ \ref{eq:Deffsim} into Eq. \ref{eq:kappa}:
\begin{equation}
\begin{split}
\frac{2}{\xi}&=\frac{\kappa}{A (\Gamma-1)}\\&=
\frac{4\eta_\mathrm{2D}(1+\xi)}{1-\eta_\mathrm{2D}(1+\xi)^2}
+\frac{2\eta_\mathrm{2D}^2(1+\xi)^3}{\left[1-\eta_\mathrm{2D}(1+\xi)^2\right]^2}.
\end{split}
\label{eq:kappaproblem}
\end{equation}
It follows that $\xi$ and thus $\bar{D}_\mathrm{eff}$ are only a function of $\eta_\mathrm{2D}$, so $f_\mathrm{pack}$ does not depend on $A$.
Instead, the parameter $A$ is only important for $f_\mathrm{or}/\eta$ through the term $2\ln{A}$.
Thus adjusting $A$ to correct the free energy is essentially equivalent to adding an extra free energy term as in previous work \cite{Graf1999}.

To solve Eq.\ \ref{eq:kappaproblem} we multiply both sides with $[1-\eta_\mathrm{2D}(1+\xi)^2]^2$ and take the leading order expression for $\xi\ll1$:
\begin{equation}
\kappa\approx 2 A (\Gamma-1)\frac{6\eta_\mathrm{2D}-5\eta_\mathrm{2D}^2}{(1-\eta_\mathrm{2D})^2}.
\label{eq:kappasim}
\end{equation}
Interestingly, it hardly matters whether one takes  Eq.\ \ref{eq:ODFsim} or the same trial function as for the nematic phase (Eq. \ref{eq:nODF}); $\bar{D}_\mathrm{eff}$ and $f_\mathrm{pack}$ have equal results and $f_\mathrm{or}/\eta$ is only increased by the additional term $1-\ln{(8/\pi)}$, which is a few percent at most for small $L/D$.
Similarly, minimizing the free energy of the nematic phase using Eq. \ref{eq:ODFsim} results in a similar expression for $f_\mathrm{pack}$ and increases $f_\mathrm{or}/\eta$ with a similar magnitude by the additional term $2\ln{(15\sqrt{\pi}/16)}-1$.
The difference between the use of a Gaussian or exponential distribution for the orientational distribution function $\psi(\theta)$ for these liquid crystal phases is therefore almost negligible.

Since $\eta_\mathrm{2D}$ is a function of $\bar{\Delta}_\bot$, a minimization of the free energy with respect to $\bar{\Delta}_\bot$ also requires simplifications in order to maintain tractable analytical expressions.
It is convenient to first use $\eta_\mathrm{2D}=\zeta\bar{\Delta}_\bot/\Gamma$, with $\zeta$ defined by:
\begin{equation}
\zeta=\frac{3\Gamma}{3\Gamma-1}\eta.
\end{equation}
Minimizing the free energy with respect to $\bar{\Delta}_\bot/\Gamma$ and taking the leading order contribution in the limit $\bar{\Delta}_\bot/\Gamma\rightarrow1$ gives:
\begin{equation}
\frac{\bar{\Delta}_\bot}{\Gamma}=1+
\frac{(6-5\zeta)(1-\zeta)^2(1-27\zeta+41\zeta^2-16\zeta^3)^2}
{\sum_{i=0}^{9} k_i \zeta^i},
\label{eq:deltasim}
\end{equation}
where the values of the constants $k_i$ are listed in Table \ref{tab:const} and a full derivation is given in Appendix \ref{sec:mindelta}.
\begin{table}
\caption{Values for $k_i$ in Eqs.\ \ref{eq:deltasim} and \ref{eq:ddeltasim}}.\label{tab:const}
\begin{ruledtabular}
\begin{tabular}{lrlrlr}
$k_0$ &12   & $k_4$ & -288,904 & $k_8$ &-105,024 \\
$k_1$  & -385 &  $k_5$ &534,956 & $k_{9}$ &14,080 \\
$k_2$  & -4,980 & $k_6$ &-553,098   \\ 
$k_3$ &    75,048& $k_7$ &328,296\\
\end{tabular}
\end{ruledtabular}
\end{table}

\subsection{AAA crystal phase}
The free energy of the AAA crystal phase can also be described by a cell theory model similar to the numerical solution of Graf and L\"{o}wen \cite{Graf1999}.
Here each spherocylinder is confined in a discrete hexagonal prism with a cross-sectional area of $\sqrt{3} \Delta_\parallel^2/2$ and height $\Delta_\bot$.
The height of these prisms $\Delta_\bot$ is similar to the lattice spacing used for the smectic-A phase.
The hexagonal prism is the Wigner--Seitz cell \cite{Wigner1933} of perfectly aligned spherocylinders in an AAA crystal.
It follows that the volume of this cell equals the available volume per particle $1/\rho$:
\begin{align}
\frac{1}{\rho}&=\frac{\sqrt{3}\Delta_\parallel^2\Delta_\bot}{2}.
\end{align}
The close-packed volume fraction of an AAA crystal, where $\Delta_\parallel\to D$ and $\Delta_\bot\to L+D$ is thus:
\begin{align}
{\eta_\mathrm{cp,AAA}}&=\frac{\pi\left(3\Gamma-1\right)}{6\sqrt{3}\Gamma},
\label{eq:cpAAA}
\end{align}
and it follows that
\begin{align}
\bar{\Delta}_\bot&=\frac{x \Gamma}{\bar{\Delta}_\parallel^2},
\label{eq:dbotAAA}
\end{align}
where $x=\eta_\mathrm{cp,AAA}/\eta$ and $\bar{\Delta}_i=\Delta_i/D$.
For aligned spherocylinders the free volume in this cell is given by:
\begin{equation}
V_{\mathrm{free,al}}=\frac{\sqrt{3}\left(\Delta_\parallel-D\right)^2\left(\Delta_\bot-L-D\right)}{2}.
\end{equation}
This leads to the following free energy for aligned spherocylinders:
\begin{equation}
\begin{split}
\frac{f_\mathrm{AAA,al}}{\eta}&=\ln\frac{\Lambda^3}{v_0}-\ln{\frac{6\sqrt{3}}{\pi\left(3\Gamma-1\right)}}\\
&\qquad-\ln{\left(\bar{\Delta}_\parallel-1\right)^2}-\ln{\left(\bar{\Delta}_\bot-\Gamma\right)},
\end{split}
\end{equation}
or using Eqs.\ \ref{eq:cpAAA} and \ref{eq:dbotAAA}:
\begin{equation}
\begin{split}
\frac{f_\mathrm{AAA,al}}{\eta}&=\ln\frac{\Lambda^3}{v_0}+\ln{\eta_\mathrm{cp,AAA}}\\
&\qquad-\ln{\left(\bar{\Delta}_\parallel-1\right)^2}-\ln{\left(\frac{x}{\bar{\Delta}_\parallel^2}-1\right)}.
\end{split}
\label{eq:fAAAal}
\end{equation}

This free energy represents the free energy of a 2D lattice of discs combined with that of a 1D lattice, representing the projections perpendicular and parallel respectively to the (fixed) direction of each rod.
To remain consistent with the treatment of the smectic-A phase, the effect of weak orientational fluctuations of the rods is estimated by substituting the free energy of the lattice of 2D discs with that of effective 2D discs with diameter $D_\mathrm{eff}$ as described in Eq.\ \ref{eq:Deff}:
\begin{equation}
\begin{split}
\frac{f_\mathrm{AAA}}{\eta}&=f_\mathrm{or}+\ln\frac{\Lambda^3}{v_0}+\ln{\eta_\mathrm{cp,AAA}}\\&\qquad-\ln{\left(\bar{\Delta}_\parallel-\bar{D}_\mathrm{eff}\right)^2}-\ln{\left(\frac{x}{\bar{\Delta}_\parallel^2}-1\right)},
\end{split}
\label{eq:fAAA}
\end{equation}
where the parameter $A=0.225\eta h$ in the definition of $D_\mathrm{eff}$ was chosen based on comparison with simulation results for the equations of state and the AAA--ABC phase transition \cite{Mcgrother1996, Bolhuis1997}.
The dependence of the parameter $A$ on the volume fraction $\eta$ again leads to an improvement in the comparison with the equations of state from the simulations.
This value of $A$ is however significantly lower than for the smectic-A phase and this is most likely due to an underestimation of the free volume from neglecting the penetration of rods into neighboring cells.

\begin{figure}
\includegraphics[width=\linewidth]{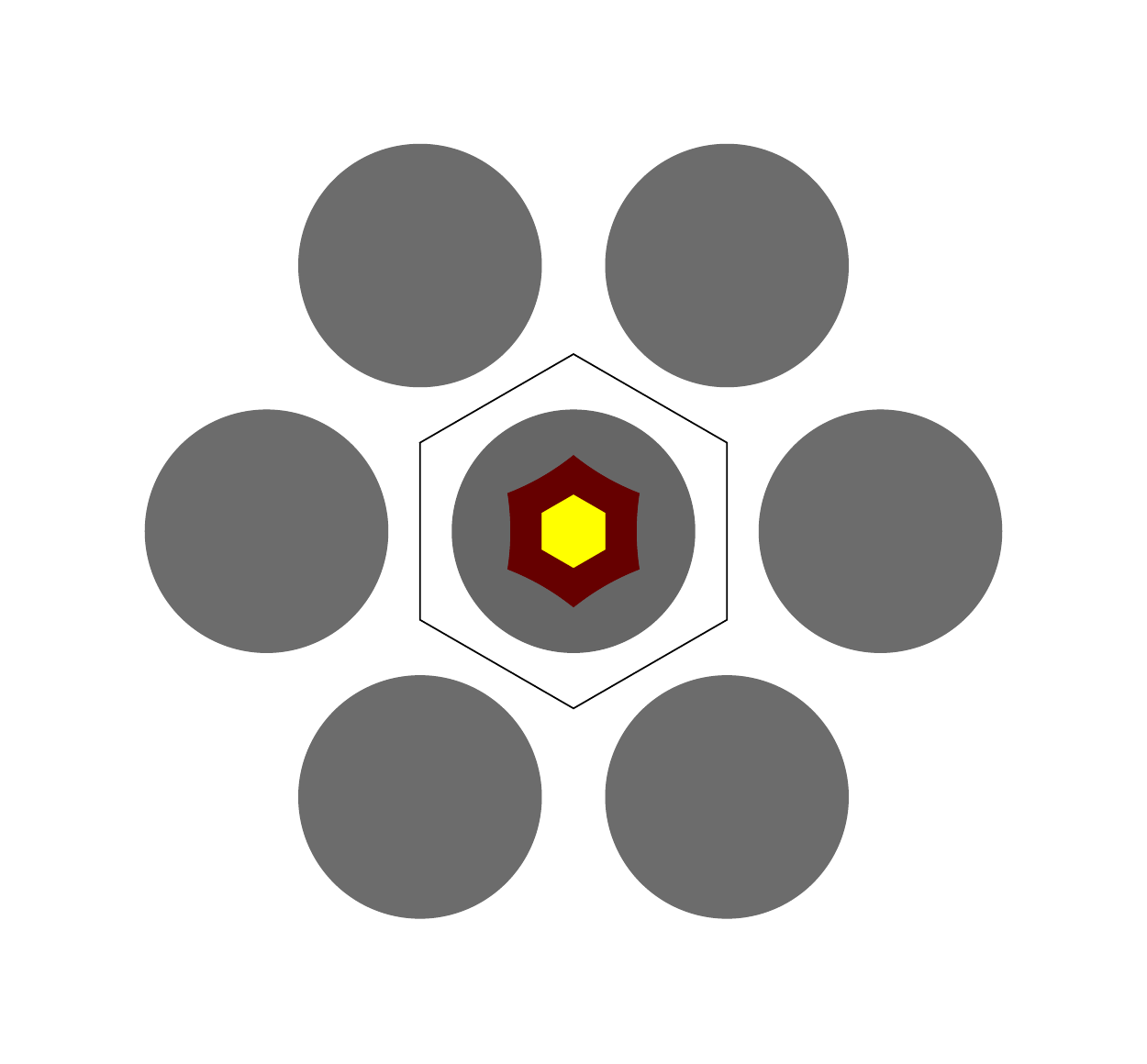}
\caption{Two types of free area in a 2D hexagonal crystal are indicated for the central particle.
The yellow area denotes the free area the particle has when confined within the Wigner--Seitz cell, which is indicated by the solid lines.
The red and yellow area combined denote the free area the particle has within the confinement of the neigboring particles.}
 \label{fig:freearea}
\end{figure}

We should note that the expressions for the free energy of disks in a 2D fluid or lattice lead to a significant deviation on the fluid--solid phase transition observed in computer simulations \cite{Doge2004,Bernard2011}.
Better agreement is obtained by using a free energy for the 2D lattice based on the free area of a particle confined by their neighboring particles which are fixed on their average position \cite{Stillinger1965,Bernard2011}.
To illustrate the difference in free area with confining the particle into a Wigner--Seitz cell, an example is given in Figure \ref{fig:freearea}.
The shape of the free area resembles a hexagon in both cases, but the length scale of this hexagon differs by a factor $2$ and thus the area by a factor $\sim4$.
This leads to an extra constant term  $-\ln{4}$ to the free energy per particle $f/\eta$.
Using a similar approach for the free length of a 1D lattice $f/\eta$ would decrease by $-\ln{2}$.
In the case of the smectic-A and AAA phase this would lead to an extra constant term  $-\ln{k}$ with $k=2$ and $k=8$ respectively.
This extra term was absorbed in the correction by the parameter $A$, but when we decouple $A$ and $k$ for the smectic-A and AAA phase, we would obtain $A=0.58\eta h$ and $A=0.64\eta h$, respectively.
The difference in the parameter $A$ is thus much smaller when the factor $k$ is considered.
For simplicity we have however only included parameter $A$ in our expressions as the free energy is equivalent and the factor $k$ can only be approximated.

\subsubsection*{Free energy minimization}
\label{sec:minFAAA}
The free energy of Eq.\ \ref{eq:fAAA} is minimized with respect to $\psi(\bold{\Omega})$ under the normalization constraint and leads to:
\begin{align}
\frac{2A\left(\Gamma-1\right)}{\bar{\Delta}_\parallel-\bar{D}_\mathrm{eff}} |\sin{\theta}|
+\log{\left[4\pi \psi(\theta)\right]}
-\lambda=0.
\end{align}
By defining a parameter $\kappa$ as:
\begin{equation}
\kappa=\frac{2A\left(\Gamma-1\right)}{\bar{\Delta}_\parallel-\bar{D}_\mathrm{eff}}
\label{eq:kappaAAA},
\end{equation}
the resulting $\psi(\bold{\theta})$, $\bar{D}_\mathrm{eff}$, and $f_\mathrm{or}$ can be approximated for $\kappa\gg 1$ similar to Eqs.\ \ref{eq:ODFsim}, \ref{eq:Deffsim}, and \ref{eq:frotsim}.
This means that also for the AAA phase the parameter $A$ is only affecting the contribution $2\ln{A}$ but not $\bar{D}_\mathrm{eff}$.
The parameter $\kappa$ is found analytically from solving Eq.\ \ref{eq:kappaAAA} using Eq.\ \ref{eq:Deffsim}:
\begin{equation}
\kappa = \frac{4A\left(\Gamma-1\right)}{\bar{\Delta}_\parallel-1}.
\label{eq:kappasimAAA}
\end{equation}
Minimizing the free energy of Eq.\ \ref{eq:fAAA} with respect to $\bar{\Delta}_\parallel$ provides the following analytical solution for $\bar{\Delta}_\parallel$:
\begin{equation}
\bar{\Delta}_\parallel =\frac{6^{1/3}x+\left(9x+x\sqrt{3\left(27-2x\right)}\right)^{2/3}}{6^{2/3}\left(9x+x\sqrt{3\left(27-2x\right)}\right)^{1/3}}.
\label{eq:deltasimAAA}
\end{equation}
In the close-packed limit ($x=1$) this expression reduces to $\bar{\Delta}_\parallel=1$ as expected.
From Eq.\ \ref{eq:dbotAAA} the value of $\bar{\Delta}_\bot$ can also be obtained.
Note that for aligned spherocylinders the expressions $\bar{\Delta}_\parallel=x^{1/3}$ and $\bar{\Delta}_\bot=\Gamma x^{1/3}$ are obtained.
This result was used by Graf and L\"{o}wen \cite{Graf1999} instead of the free energy minimization with respect to the lattice constants.

\subsection{ABC crystal phase}
To describe the free energy of the ABC crystal Graf and L\"{o}wen also used cell theory \cite{Graf1999}.
Here, the free volume is assumed to be shaped as a rhombic dodecahedron analogous to the corresponding fcc crystal of hard spheres.
In making this assumption the distances between the particles are fixed in a position that does not necessarily correspond to the free energy minimum.
Thus we base our approach on the cell theory result of Taylor \textit{et al.} for aligned spherocylinders \cite{Taylor1989}.
In this theory each spherocylinder is confined to a discrete hexagonal tube with a cross-sectional area of $\sqrt{3} \Delta_\parallel^2/2$ and height $\Delta_\bot$.
The ends of the tube are capped with hemi-dodecahedrons of total volume $\sqrt{2} \Delta_\parallel^3/2$. 
The shape of these caps is based on the shape of the Wigner--Seitz cell of an fcc crystal for spheres, which is a rhombic dodecahedron.
The cell is similar to that of the AAA crystal except at the ends.
The height $\Delta_\bot$ of the hexagonal tube is therefore smaller than $\Delta_\bot$ for the AAA or smectic-A phase.
As these cells should be close-packed and space-filling, the total volume of the cell is equal to $1/\rho$ :
\begin{align}
\frac{1}{\rho}&=\Delta_{\parallel,\mathrm{al}}^2\frac{\sqrt{3}\Delta_{\bot,\mathrm{al}}+\sqrt{2}{\Delta_{\parallel,\mathrm{al}}}}{2}.
\end{align}

In the limit of $\Delta_{\parallel,\mathrm{al}}\to D$ and $\Delta_{\bot,\mathrm{al}}\to L$ this leads to close-packing of hard spherocylinders:
\begin{align}
\eta_\mathrm{cp}&=\frac{\pi\left(3\Gamma-1\right)}{6\left(\sqrt{3}\left(\Gamma-1\right)+\sqrt{2}\right)}.
\label{eq:cpABC}
\end{align}
The expression for $\bar{\Delta}_{\bot,\mathrm{al}}$ can be written as:
\begin{align}
\bar{\Delta}_{\bot,\mathrm{al}}&=\frac{x_{\mathrm{al}}}{\bar{\Delta}_{\parallel,\mathrm{al}}^2}\left(\left(\Gamma-1\right)+\frac{\sqrt{2}}{\sqrt{3}}\right)-\frac{\sqrt{2}\bar{\Delta}_{\parallel,\mathrm{al}}}{\sqrt{3}},
\label{eq:dbotABCal}
\end{align}
where $x_\mathrm{al}=\eta_\mathrm{cp}/\eta$.
The free volume of the aligned spherocylinders inside this cell is given by:
\begin{equation}
\begin{split}
&V_{\mathrm{free,al}}=\\&\left(\Delta_{\parallel,\mathrm{al}}-D\right)^2\frac{\sqrt{3}\left(\Delta_{\bot,\mathrm{al}}-L\right)+\sqrt{2}\left(\Delta_{\parallel,\mathrm{al}}-D\right)}{2}.
\end{split}
\end{equation}
The free energy then becomes:
\begin{equation}
\begin{split}
&\frac{f_\mathrm{ABC,al}}{\eta}=\ln\frac{\Lambda^3}{v_0}-\ln{\frac{6\sqrt{3}}{\pi\left(3\Gamma-1\right)}}-\ln{\left(\bar{\Delta}_{\parallel,\mathrm{al}}-1\right)^2}\\
&-\ln{\left(\left(\bar{\Delta}_{\bot,\mathrm{al}}-\left(\Gamma-1\right)\right)+\frac{\sqrt{2}}{\sqrt{3}}\left(\bar{\Delta}_{\parallel,\mathrm{al}}-1\right)\right)}.
\end{split}
\end{equation}
Using Eqs.\ \ref{eq:cpABC} and \ref{eq:dbotABCal} this can be rewritten as:
\begin{equation}
\begin{split}
\frac{f_\mathrm{ABC,al}}{\eta}&=\ln\frac{\Lambda^3}{v_0}+\ln{\eta_\mathrm{cp}}\\
&-\ln{\left(\bar{\Delta}_{\parallel,\mathrm{al}}-1\right)^2}
-\ln{\left(\frac{x_\mathrm{al}}{\bar{\Delta}_{\parallel,\mathrm{al}}^2}-1\right)}.
\end{split}
\end{equation}
Notice the strong similarity with Eq.\ \ref{eq:fAAAal}, where the only difference is in the close-packed volume fraction.

To include orientational fluctuations a similar approach can again be used by replacing the free energy of a 2D disc hexagonal crystal with that of an effective 2D disc with diameter $D_\mathrm{eff}$.
The subsequent free energy minimization is the same as for the AAA crystal with the only difference being the filling fraction at close-packing.
While this suffices to reproduce the equation of state and the phase coexistence curves obtained from computer simulations \cite{Mcgrother1996,Bolhuis1997}, the agreement could be improved even further.
For the previously discussed phases discrepancies in the free volume were corrected by the parameter $A$ (or $k$), but for the ABC phase the discrepancy in the tube free volume and the end caps is not necessarily the same and thus a single parameter might be insufficient.
Instead we assume the free volume to take the following form:
\begin{equation}
\begin{split}
&V_{\mathrm{free}}=\\
&\left(\Delta_{\parallel}-D_\mathrm{eff}\right)^2\frac{\sqrt{3}\left(\Delta_{\bot}-L\right)+\sqrt{2}B\left(\Delta_{\parallel}-D\right)}{2},
\end{split}
\end{equation}
where we have introduced a second correction parameter $B$, which is assumed constant.
The relation between $\Delta_{\bot}$ and $\Delta_{\parallel}$ is now approximated as:
\begin{align}
\bar{\Delta}_{\bot}&\approx\frac{x}{\bar{\Delta}_{\parallel}^2}\left(\left(\Gamma-1\right)+\frac{\sqrt{2}B}{\sqrt{3}}\right)-\frac{\sqrt{2}B\bar{\Delta}_{\parallel}}{\sqrt{3}},
\label{eq:dbotABC}
\end{align}
where $x=\eta_\mathrm{ref}/\eta$ and
\begin{align}
\eta_\mathrm{ref}&=\frac{\pi\left(3\Gamma-1\right)}{6\left(\sqrt{3}\left(\Gamma-1\right)+\sqrt{2}B\right)}.
\label{eq:crABC}
\end{align}
The free energy of the ABC crystal is then given by
\begin{equation}
\begin{split}
\frac{f_\mathrm{ABC}}{\eta}&=f_\mathrm{or}+\ln\frac{\Lambda^3}{v_0}+\ln{\eta_\mathrm{ref}}\\
&-\ln{\left(\bar{\Delta}_\parallel-\bar{D}_\mathrm{eff}\right)^2}-\ln{\left(\frac{x}{\bar{\Delta}_\parallel^2}-1\right)},
\end{split}
\label{eq:fABC}
\end{equation}
with parameters $A=0.239\eta h$ and $B=1.16$.
Here the parameter $B$ was chosen primarily to match with the osmotic pressure data from the simulations of McGrother \textit{et al.} \cite{Mcgrother1996} which is unaffected by the interpolations.
The parameter $A$ was chosen to match the ABC--smectic-A phase coexistence simulation results \cite{Bolhuis1997}, while retaining the same $\eta$ dependence as in the other phases.
Notice that the value of $A$ is quite similar to that of the AAA crystal.
After minimizing the free energy $D_\mathrm{eff}$, $f_\mathrm{or}$, and $\bar{\Delta}_\parallel$ are given by Eqs.\ \ref{eq:Deffsim}, \ref{eq:frotsim}, \ref{eq:deltasimAAA} with $x=\eta_\mathrm{ref}/\eta$ and $\kappa$ follows from Eq.\ \ref{eq:kappasimAAA}.
The expressions, however, break down near the close-packing since the parallel spacing becomes smaller than the spherocylinder length which is unphysical.

\subsection{Phase behavior; binodals}
The free energy expressions can be applied to predict the phase behavior of hard spherocylinders.
We calculate the concentrations at the binodal by solving the coexistence equations for the phases $\RN{1}$ and $\RN{2}$:
\begin{align}
\widetilde{\mu}_{\RN{1}}&=\widetilde{\mu}_{\RN{2}},
\label{eq:co1}\\
\widetilde{\Pi}_{\RN{1}}&=\widetilde{\Pi}_{\RN{2}}.
\label{eq:co2}
\end{align}
Here the normalized chemical potential $\widetilde{\mu}=\mu/(k_\mathrm{B}T)$ follows from $\widetilde{\mu}=\partial f/\partial\eta$ and the normalized pressure $\widetilde{\Pi}=\Pi v_0/(k_\mathrm{B}T)$ follows from $\widetilde{\Pi}=\eta\widetilde{\mu}-f$.
Using Eqs.\ \ref{eq:intftot}, \ref{eq:FtotalSPT}, \ref{eq:fAAA}, and \ref{eq:fABC} the expressions for $\widetilde{\mu}$ and $\widetilde{\Pi}$ of the isotropic, nematic, smectic-A, AAA and ABC phase are given in Appendix \ref{sec:expin}, \ref{sec:exp}, and \ref{sec:expcr}.
By solving Eqs.\ \ref{eq:co1} and \ref{eq:co2} for two of these phases, the concentrations at the binodal of these two phases is obtained.
At a specific $L/D$ and $\eta$ the single phase or phase coexistence with the lowest $f$ is the most stable.
For certain $L/D$ it is also possible to have coexistence with a third phase, where $\widetilde{\mu}_{\RN{1}}=\widetilde{\mu}_{\RN{3}}$ and $\widetilde{\Pi}_{\RN{1}}=\widetilde{\Pi}_{\RN{3}}$ holds in addition to Eqs. \ref{eq:co1} and \ref{eq:co2}.

\section{Results and Discussion}
\begin{figure}
\includegraphics[width=\linewidth]{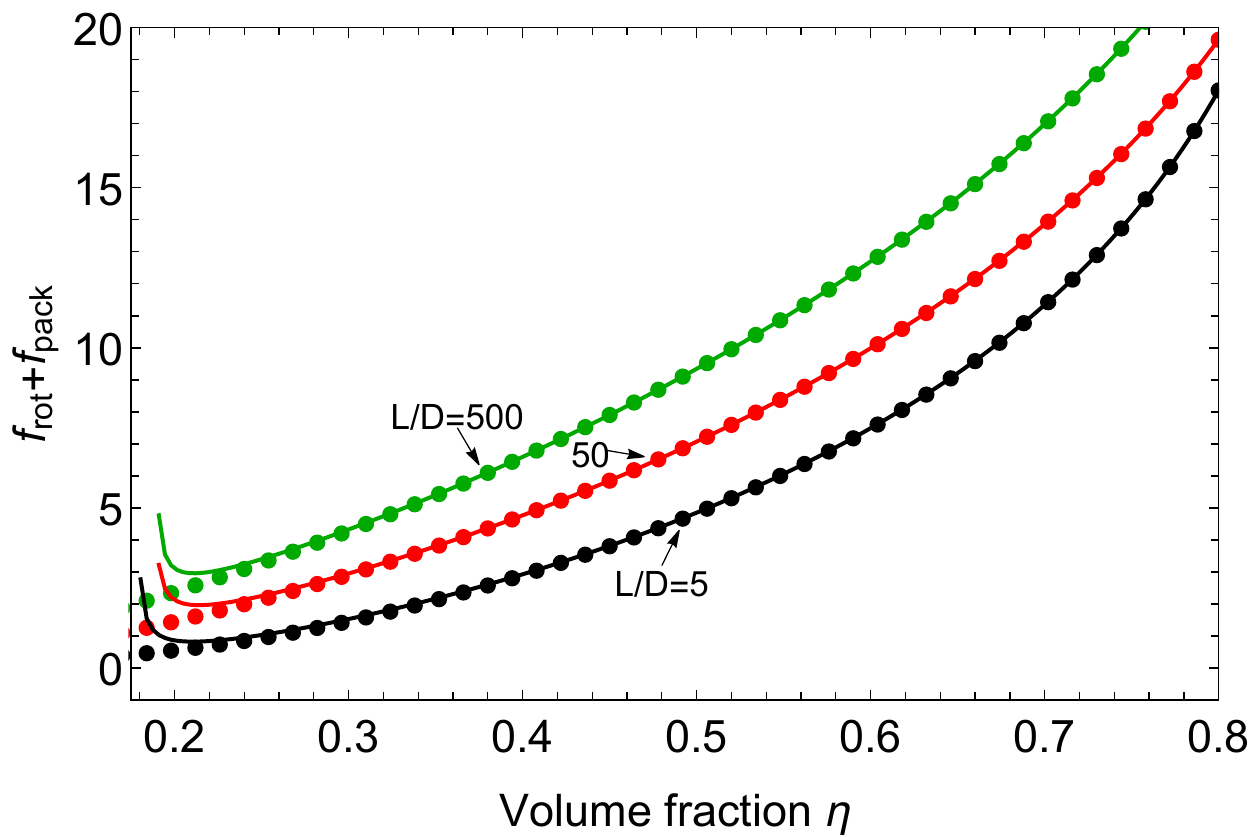}
\caption{Volume fraction dependence of the excess free energy of the smectic-A phase of Eq.\ \ref{eq:FtotalSPT}, $f_\mathrm{or}+f_\mathrm{pack}$.
Free energy obtained using numerical minimization (data points) is compared to approximate analytical results (solid curves) for aspect ratios $L/D=5$, $50$ and $500$, where both use the correction parameter $A=0.41\eta h$ based on fits to the simulation data \cite{Mcgrother1996, Bolhuis1997}.}
 \label{fig:fsmA}
\end{figure}

Here we will study the accuracy of the new analytical free energy expressions for the smectic-A, AAA crystal and ABC crystal phases in predicting the phase behavior of hard spherocylinders.
We put less emphasis on the isotropic--nematic phase transition as this has already been examined more in depth using both PL and SPT theory \cite{Franco-Melgar2008,Tuinier2016}.
First we focus on verifying the analytical free energy of the smectic-A phase by comparing it to numerical results.
The predicted excess free energy, $f_\mathrm{or}+f_\mathrm{pack}$, of the smectic-A phase for aspect ratios $L/D=5$, $50$ and $500$ is plotted in Figure \ref{fig:fsmA} for both numerical minimization of Eq.\ \ref{eq:FtotalSPT} (data points) and the simplified analytical minimization of section \ref{sec:minF} (solid curves).
It is clear that the difference between the simplified analytical and numerical minimization is negligible for $\eta \gtrsim 0.25 $ in all cases.
At lower volume fractions the assumption of $\kappa\gg1$ and the analytical expression break down due to relatively large orientational freedom for dilute rods.
The smectic-A phase is however metastable at those rod concentrations \cite{Mcgrother1996,Bolhuis1997}, so this deviation is irrelevant for our purpose.
The same arguments hold for the crystal phases.

\begin{figure*}
\subfloat{%
\resizebox*{0.5\linewidth}{!}{\includegraphics{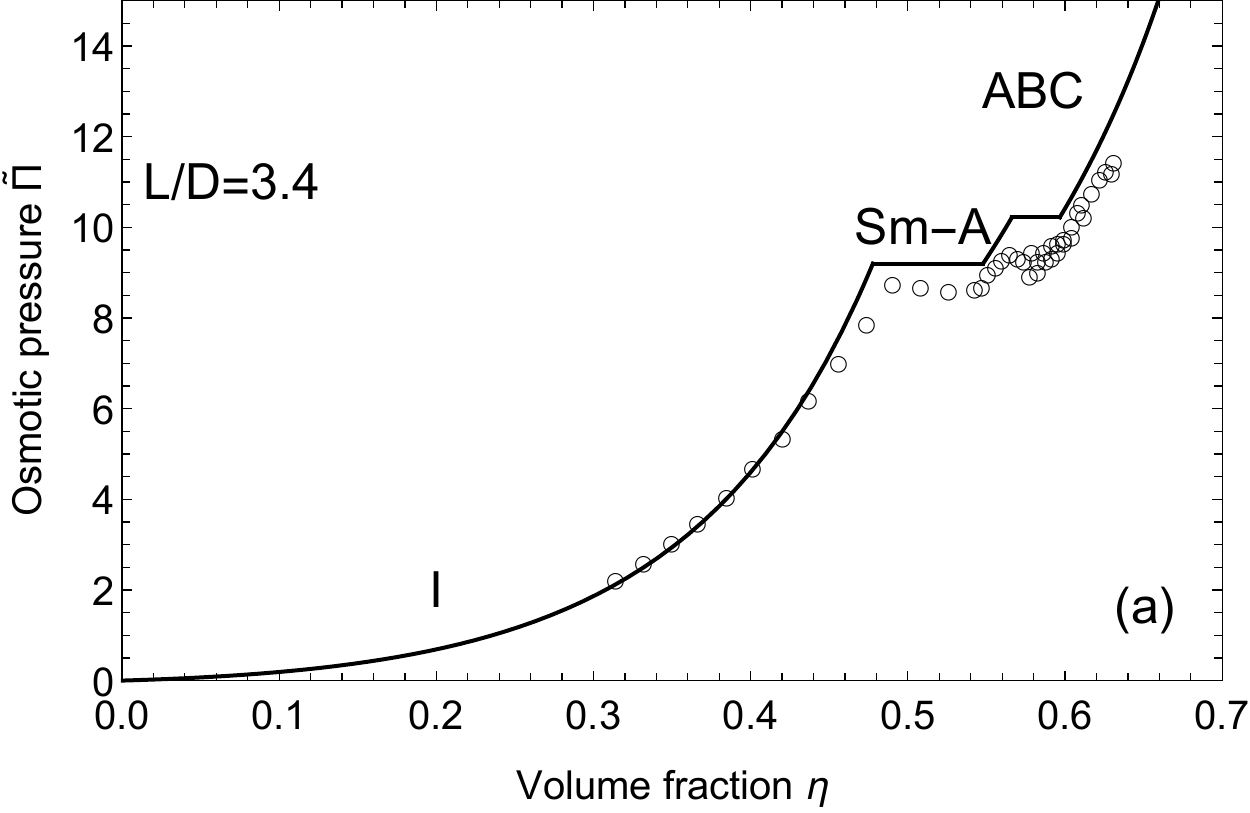}}}
\subfloat{%
\resizebox*{0.5\linewidth}{!}{\includegraphics{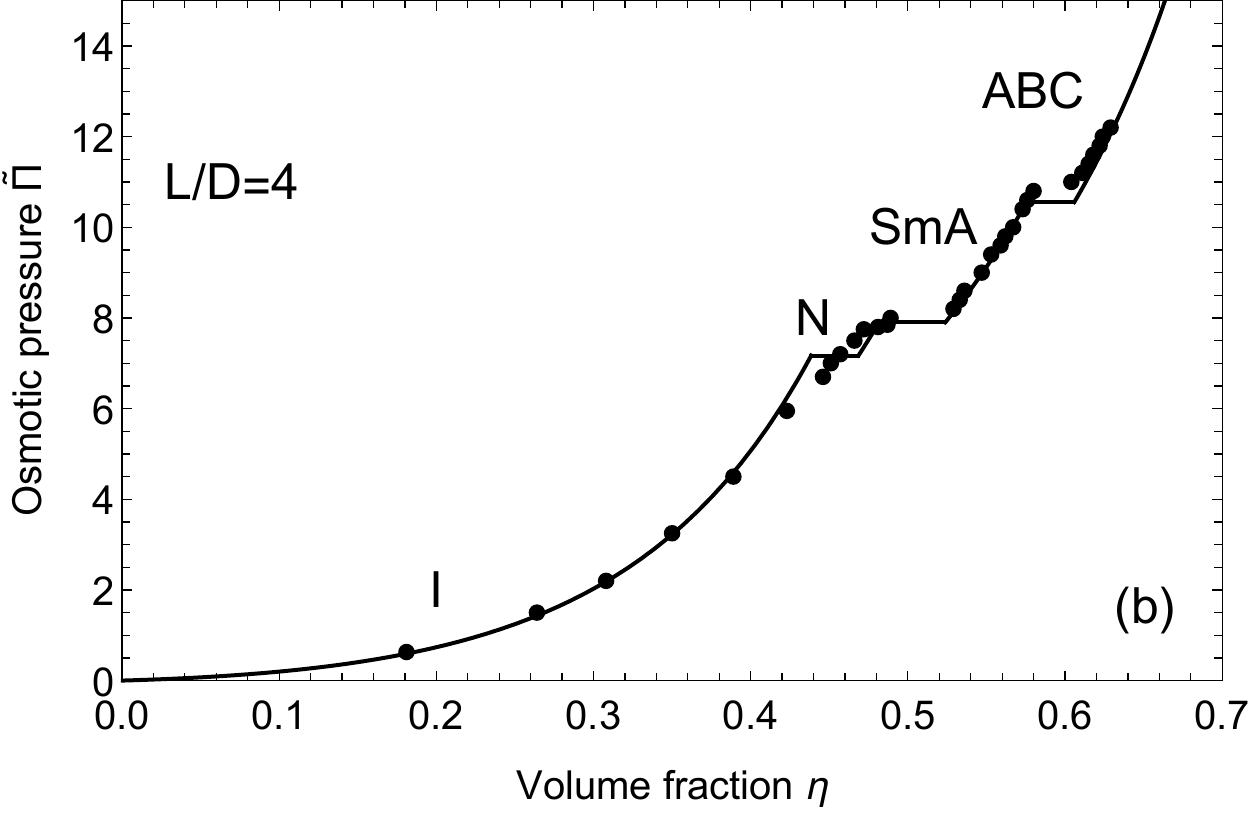}}}\\
\subfloat{%
\resizebox*{0.5\linewidth}{!}{\includegraphics{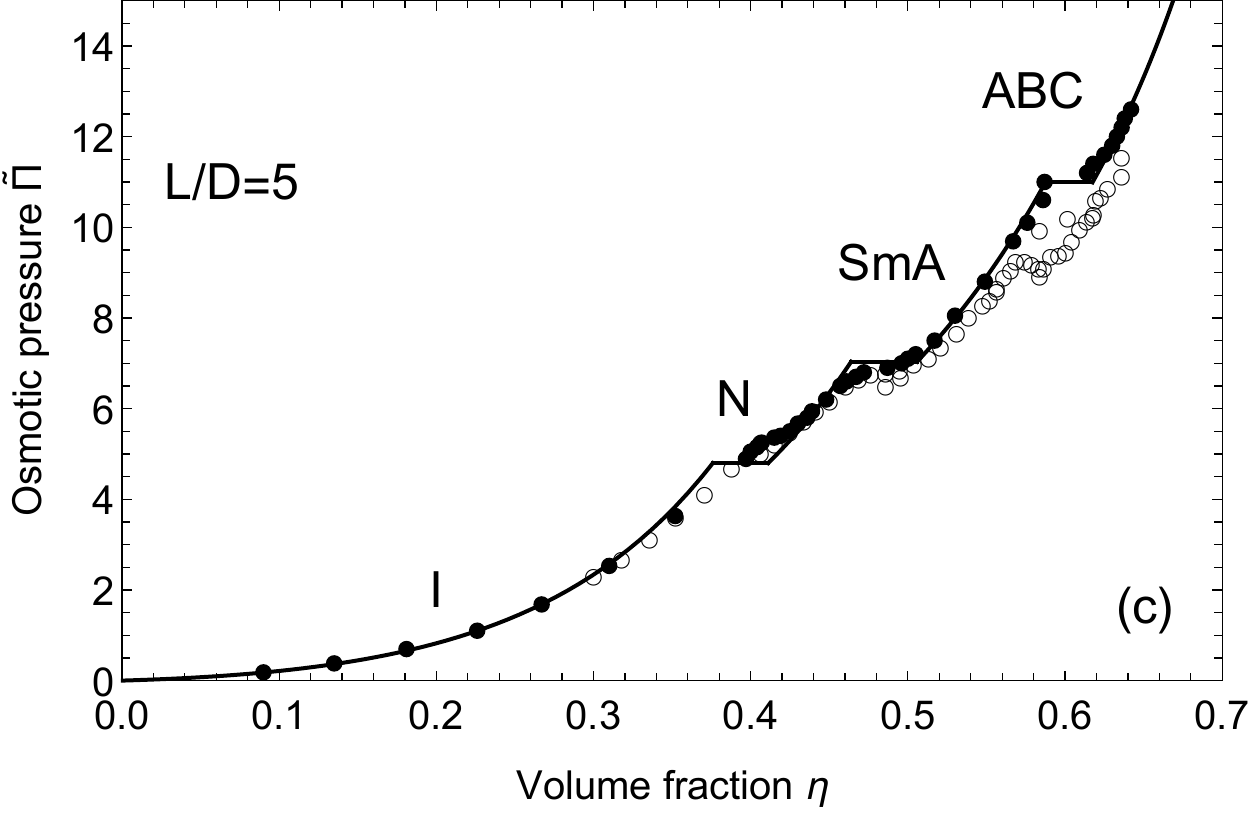}}}
\subfloat{%
\resizebox*{0.5\linewidth}{!}{\includegraphics{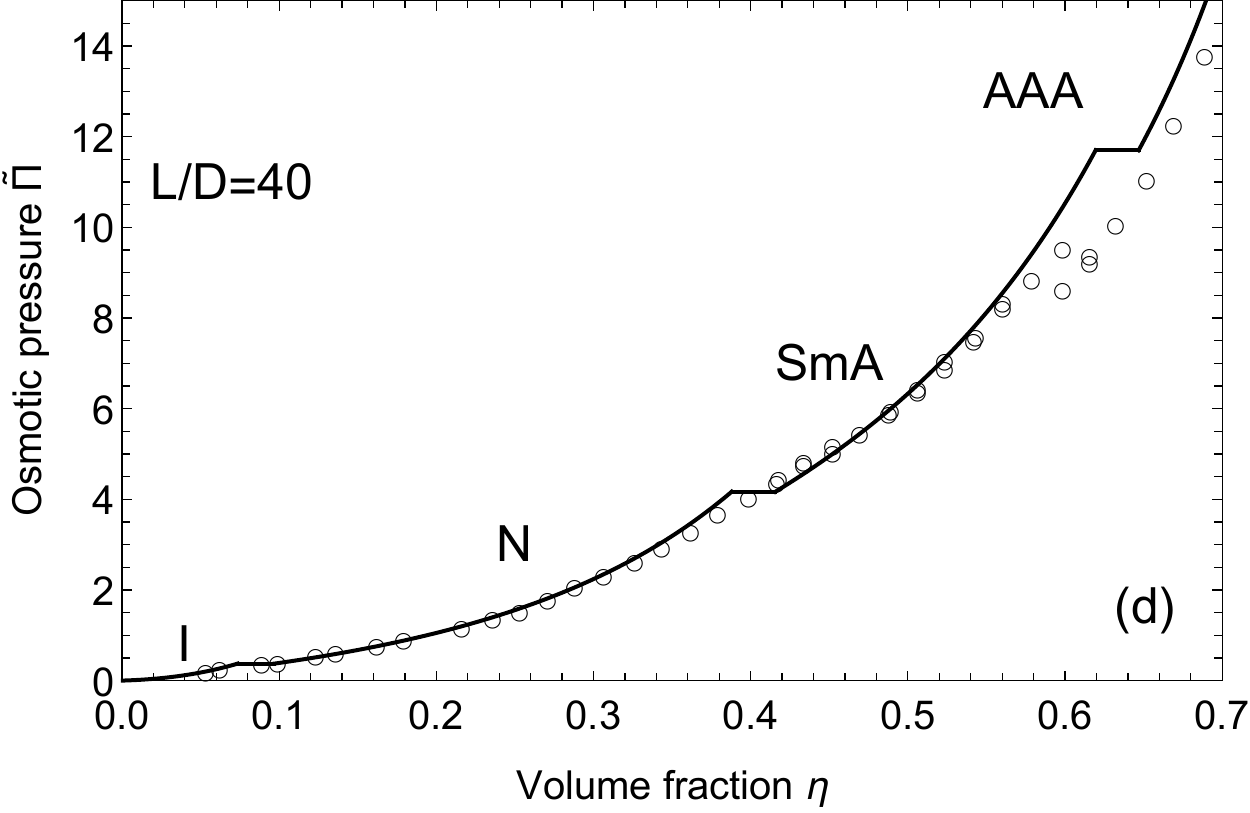}}}
\caption{Predicted equations of state (curves) for hard spherocylinders of aspect ratio $L/D=$ (a) 3.4, (b) 4, (c) 5, and (d) 40 as indicated.
The symbols are simulation results from McGrother \textit{et al.} \cite{Mcgrother1996} (filled) and from Bolhuis and Frenkel \cite{Bolhuis1997} (open).} \label{fig:eos}
\end{figure*}

Next we show in Figure \ref{fig:eos} a comparison of our analytical equations of state with simulation results for various $L/D$ values  \cite{Mcgrother1996,Bolhuis1997} (see Eqs.\ \ref{eq:pin}, \ref{eq:psm}, and \ref{eq:pcrex} for the explicit equations for the osmotic pressure).
For the smectic-A and ABC phase we find good agreement for all studied $L/D$ values in comparison to the simulation results of McGrother \textit{et al.} \cite{Mcgrother1996}
The simulation results of Bolhuis and Frenkel \cite{Bolhuis1997} show a slightly lower osmotic pressure for the smectic-A and crystal phases.
The inclusion of $\eta$ in $A$ leads to an additional term in $\widetilde{\Pi}$ of $2\eta$, which is the approximate difference between the uncorrected pressure of the smectic-A phase and the simulation results for all $L/D$ values.
The agreement with simulations is a strong improvement with respect to the reported equations of state from density functional theory \cite{Wittmann2014}.

\begin{figure}
\includegraphics[width=\linewidth]{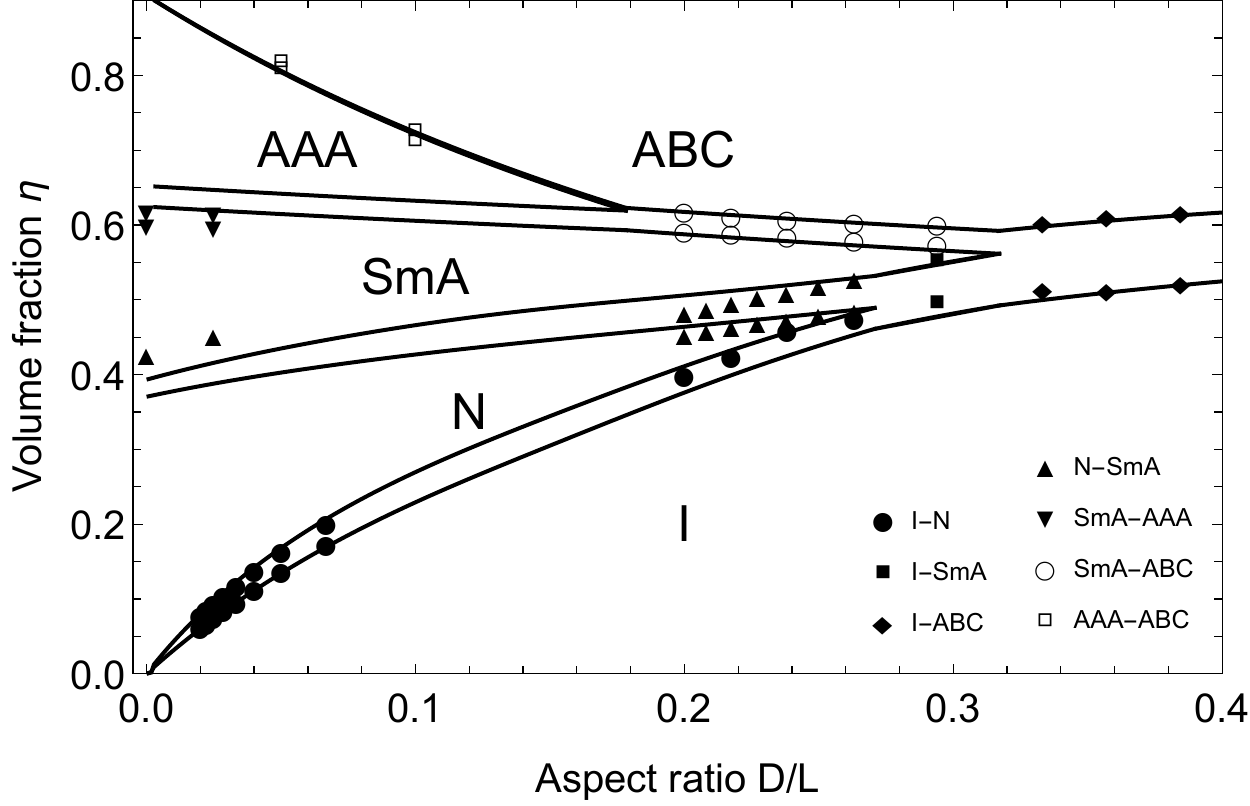}
\caption{Phase diagram of hard spherocylinders in terms of the volume fraction $\eta$ and inverse aspect ratio $D/L$.
Curves are our predictions based on algebraic expressions for all phase states.
The data points are simulation results from Bolhuis and Frenkel \cite{Bolhuis1997}.}
 \label{fig:phase}
\end{figure}

The phase coexistences resulting the analytical equations of state are as plotted (curves) in Figure \ref{fig:phase} as a function of the inverse aspect ratio $D/L$.
The phase behavior is compared to Monte Carlo simulation results (data points) for hard spherocylinders \cite{Bolhuis1997}.
A particular region of interest in this work is the nematic--smectic-A coexistence, which is in reasonable agreement with the simulations.
The coexistence drops slightly to lower volume fractions as $D/L$ is decreased.
For this coexistence the best agreement with the simulations was achieved for short rods using PL for the nematic phase and $A=0.28\eta$ as correction parameter for the smectic-A phase.
Due to inconsistencies between the two theories however, the transition drops to volume fractions below 0.3 for $D/L\lesssim0.1$.
Using SPT for the nematic phase the best agreement for short rods was achieved with $A=0.41\eta$, which still retains a similar concentration range for the phase transition of long rods.
The main issue with SPT however is that it leads to different triple points for short rods: the nematic phase is predicted to become stable at higher $D/L$ than the smectic-A phase.
Thus while the SPT free energy expressions provides a better general description, it is more appropriate to use the Parsons--Lee description for the short rod region.
This has led to our use of the aforementioned sigmoidal interpolation between the two equations of state.
Note that a higher value of $A$ would shift the phase coexistence to higher volume fractions for all $L/D$.
Without any correction, the volume fraction of the nematic--smectic-A coexistence curves would increase by about 0.1--0.2.
The choice of parameter $A$ for the ABC and AAA phase gives excellent agreement with simulations for the smectic-A--ABC and AAA--ABC phase coexistence.
Additionally, the smectic-A--AAA, isotropic--smectic-A and isotropic--ABC  phase coexistence conforms to computer simulations, while these did not directly influence the choices for $A$.

Comparing the results in Figure \ref{fig:phase} to the previous numerical results by Graf and L\"{o}wen \cite{Graf1999}, the agreement with simulations of all coexistence curves is improved.
The main reason for the improvement of the nematic--smectic-A coexistence curve comes from a more accurate choice for the nematic free energy and the inclusion of $\eta$ in the correction term.
For both crystal phases the improvement comes from the free energy minimization over the lattice constants and the different corrections in $A$ and $B$.
The most recent density functional theories on the nematic--smectic-A phase transition \cite{Velasco2000,Wittmann2014,Wittmann2016} have shown improved or comparable agreement for short rods with simulations.
The main deviation between our results and density functional theory results is in the order of the nematic--smectic-A transition at small $D/L$.
While the order of the transition could not be determined conclusively in simulations \cite{Bolhuis1997}, it is implied in our method that all phase transitions represent discontinuous first order phase transitions.
Density functional theory however predicts for long rods that the nematic--smectic-A phase transition becomes a continuous second order phase transition after a certain tricritical point \cite{Poniewierski1990,Somoza1990,Wittmann2014,Wittmann2016}, but the location of this point is unclear.
Similar to our results in the Onsager limit ($D/L\to0$) the bifurcation point of this second order phase transition is predicted to be near $\eta\sim0.4$ \cite{Somoza1990,Wittmann2016,Poniewierski1992}, which is slightly below the value found in the simulations.
Interestingly, simulations performed for semi-flexible hard spherocylinders revealed this phase transition to be first order \cite{DeBraaf2017}.

In addition density functional theory, and in particular fundamental measure theory, has been extended to include arbitrary convex particle shapes, though it remains numerically involved \cite{Hansen-Goos2009,Wittmann2015}.
As our focus has been on deriving algebraic free energy expressions for hard spherocylinders in particular, the presented expressions are not applicable for different particle shapes.
For similar uniaxial and convex shaped particles as regular hard cylinders \cite{Wensink2009} or hexagonal plates it should however be possible to use our methodology to derive new algebraic free energy expressions by adjusting the geometrical considerations behind the excluded volume (fluid phases) and cell free volume (crystalline phases).


\section{Concluding Remarks}
We have presented a comprehensive algebraic description for the free energies of the principal thermodynamic phases (including the smectic-A, AAA, and ABC phase) of hard rod-like particles providing expressions for the equations of state.
Based on previous numerical calculations, these expressions provide a computationally straightforward method to predict phase coexistences, while providing additional structural information for the smectic-A and crystal phases in terms of the equilibrium lattice spacings.
Using empirical corrections for the nematic, smectic-A, AAA, and ABC free energies, we find that the predicted phase behavior is in quantitative agreement with results from computer simulations and density functional theory.
An important advantage of the algebraic free energy expressions is that they pave the way towards more realistic descriptions of colloidal liquid crystals based on perturbation or free volume theories in which the effects of soft rod-rod interactions (generated by e.g. van der Waals, depletion or electrostatic forces) can be incorporated.
This is particularly relevant for understanding the role of rod flexibility and soft interactions in driving the competitive stability of smectic, columnar and crystal order in suspensions of rod-shaped colloids, which remains an outstanding issue. 

\begin{acknowledgments}
We would like to thank {\'{A}}.\ Gonz{\'{a}}lez Garc{\'{i}}a and J. Opdam for useful discussions on cell theory and the columnar phase of hard plates.
We are indebted to H.\ N.\ W.\ Lekkerkerker for useful discussions and suggestions.
MV acknowledges the Netherlands Organization for Scientific Research (NWO) for a Veni grant (no. 722.017.005).
\end{acknowledgments}

\appendix

\section{Minimization of lattice spacing}
\label{sec:mindelta}
For the minimization of the free energy with respect to $\bar{\Delta}_\bot/\Gamma$ it is convenient to first express the derivative of $f/\eta$ with respect to $\eta_\mathrm{2D}$:
\begin{equation}
\begin{split}
&\eta_\mathrm{2D}\frac{\partial f/\eta}{\partial \eta_\mathrm{2D}}=\frac{2(6-4\eta_\mathrm{2D})}{(6-5\eta_\mathrm{2D})(1-\eta_\mathrm{2D})}\\
&+\left(\frac{1}{\left(1-\eta_\mathrm{2D}\bar{D}_\mathrm{eff}^2\right)^2}-1\right)
\left(1-4\frac{(1-\eta_\mathrm{2D})(3-2\eta_\mathrm{2D})}{(6-5\eta_\mathrm{2D})^2\eta_\mathrm{2D}\bar{D}_\mathrm{eff}}\right).\\
\end{split}
\label{eq:der2D}
\end{equation}
This allows us to write the minimization condition as:
\begin{equation}
\frac{\partial f/\eta}{\partial \bar{\Delta}_\bot/\Gamma}=
\left(\eta_\mathrm{2D}\frac{\mathrm{d}f/\eta}{\mathrm{d}\eta_\mathrm{2D}}
+\frac{1}{1-\bar{\Delta}_\bot/\Gamma}\right)\frac{\Gamma}{\bar{\Delta}_\bot}=0.
\label{eq:cond}
\end{equation}
The left-hand side of Eq.\ \ref{eq:cond} can be rewritten as a polynomial of $\bar{\Delta}_\bot/\Gamma$ by multiplying with a factor:
\begin{equation}
\begin{split}
&\frac{\bar{\Delta}_\bot^2}{\Gamma^2}\left(1-\frac{\bar{\Delta}_\bot}{\Gamma}\right)
\left(6-5\eta_\mathrm{2D}\right)\left(1-\eta_\mathrm{2D}\right)^2\\
&\times\left(1-27\eta_\mathrm{2D}+41\eta_\mathrm{2D}^2-16\eta_\mathrm{2D}^3\right)^2.
\end{split}
\end{equation}
Taking the limit $\bar{\Delta}_\bot/\Gamma\to1$ of the resulting polynomial leads to a linear relation of $\bar{\Delta}_\bot/\Gamma$, which is used to find the approximate solution given by Eq.\ \ref{eq:deltasim}.

\section{Chemical potential and osmotic pressure of the isotropic and nematic phase}
\label{sec:expin}
The normalized chemical potential $\widetilde{\mu}$ and the normalized pressure $\widetilde{\Pi}$ are given by $\widetilde{\mu}=\partial f/\partial\eta$ and $\widetilde{\Pi}=\eta\widetilde{\mu}-f$.
Using the free energy expressions of Eqs.\ \ref{eq:iftot} and \ref{eq:nftot}, which follow $\partial f/\partial\kappa=0$, this gives for both the isotropic and nematic phase \cite{Tuinier2016}:
\begin{align}
\widetilde{\mu}_\mathrm{SPT}&=\frac{f_\mathrm{SPT}}{\eta}+\frac{1}{1-\eta}+a\frac{\eta}{\left(1-\eta\right)^2}+b\frac{\eta^2}{\left(1-\eta\right)^3},\\
\widetilde{\mu}_\mathrm{PL}&=\frac{f_\mathrm{PL}}{\eta}+1+\frac{2 n-n^2}{2 (1-n)^3}\left(4+\frac{3 (\Gamma -1)^2}{3 \Gamma -1}\right),\\
\widetilde{\mu}&=g \widetilde{\mu}_\mathrm{SPT}+(1-g)\widetilde{\mu}_\mathrm{PL},
\label{eq:min}\\
\frac{\widetilde{\Pi}_\mathrm{SPT}}{\eta}&=\frac{1}{1-\eta}+a\frac{\eta}{\left(1-\eta\right)^2}+b\frac{\eta^2}{\left(1-\eta\right)^3},\\
\frac{\widetilde{\Pi}_\mathrm{PL}}{\eta}&=1+\frac{2 n-n^2}{2 (1-n)^3}\left(4+\frac{3 (\Gamma -1)^2}{3 \Gamma -1}\right),\\
\widetilde{\Pi}&=g \widetilde{\Pi}_\mathrm{SPT}+(1-g)\widetilde{\Pi}_\mathrm{PL}.
\label{eq:pin}
\end{align}

\section{Chemical potential and osmotic pressure of the smectic-A phase}
\label{sec:exp}
The normalized chemical potential $\widetilde{\mu}$ and the normalized pressure $\widetilde{\Pi}$ were calculated by $\widetilde{\mu}=\partial f/\partial\eta$ and $\widetilde{\Pi}=\eta\widetilde{\mu}-f$ using Eqs.\ \ref{eq:FtotalSPT} and \ref{eq:der2D}:
\begin{align}
\begin{split}
\widetilde{\mu}&=\frac{f}{\eta}+1+\frac{2\eta A'}{A}\\
&+\eta_\mathrm{2D}\frac{\mathrm{d}f/\eta}{\mathrm{d}\eta_\mathrm{2D}}\left(1+\frac{\zeta\bar{\Delta}_\bot'}{\bar{\Delta}_\bot}\right)
+\frac{1}{1-\bar{\Delta}_\bot/\Gamma}\frac{\zeta\bar{\Delta}_\bot'}{\bar{\Delta}_\bot},
\end{split}
\label{eq:msm}
\\\begin{split}
\frac{\widetilde{\Pi}}{\eta}&=1+\frac{2\eta A'}{A}\\
&+\eta_\mathrm{2D}\frac{\mathrm{d}f/\eta}{\mathrm{d}\eta_\mathrm{2D}}\left(1+\frac{\zeta\bar{\Delta}_\bot'}{\bar{\Delta}_\bot}\right)
+\frac{1}{1-\bar{\Delta}_\bot/\Gamma}\frac{\zeta\bar{\Delta}_\bot'}{\bar{\Delta}_\bot},
\end{split}
\label{eq:psm}
\end{align}
where $\bar{\Delta}_\bot'$ is given by:
\begin{equation}
\frac{\bar{\Delta}_\bot'}{\Gamma}=\frac{-2(1-\zeta)(1-27\zeta+41\zeta^2-16\zeta^3)}
{\left(\sum_{i=0}^{9} k_i \zeta^i\right)^2}\sum_{i=0}^{12} l_i \zeta^i.
\label{eq:ddeltasim}
\end{equation}
Here the values of the constants $k_i$ and $l_i$ is given in Tables \ref{tab:const} and \ref{tab:constl}.
While these equations were used for our results, it is insightful to also give the leading order expressions, which are exact for $\partial f/\partial\kappa=0$ and $\partial f/\partial\Delta_\bot=0$:
\begin{align}
\widetilde{\mu}&=\frac{f}{\eta}+\frac{2\eta A'}{A}
+\frac{1}{\left(1-\eta_\mathrm{2D}\bar{D}_\mathrm{eff}^2\right)^2},
\label{eq:msmex}
\\
\begin{split}
\frac{\widetilde{\Pi}}{\eta}&=\frac{2\eta A'}{A}+\frac{1}{\left(1-\eta_\mathrm{2D}\bar{D}_\mathrm{eff}^2\right)^2}.
\label{eq:psmex}
\end{split}
\end{align}
These leading order equations have a maximum deviation of around $6\%$ for $\eta\geq 0.4$ and $L/D\geq3$ with Eqs.\ \ref{eq:msm} and \ref{eq:psm}. 
\begin{table}
\caption{Values for $l_i$ in Eq.\ \ref{eq:ddeltasim}}.\label{tab:constl}
\begin{ruledtabular}
\begin{tabular}{lrlrlr}
 $l_{0}$ &891 &  $l_{5}$ &-101,462,282& $l_{10}$ &27,625,680\\
$l_{1}$ &-74,532& $l_{6}$& 173,571,605 & $l_{11}$ &-5,499,680 \\
$l_{2}$ &999,914 & $l_{7}$&-198,678,876& $l_{12}$ &487,680\\ 
$l_{3}$ &-8,373,233   & $l_{8}$&154,739,944 \\
 $l_{4}$ &37,933,935 & $l_{9}$& -81,271,047 \\
\end{tabular}
\end{ruledtabular}
\end{table}

\section{Chemical potential and osmotic pressure of the crystal phases}
\label{sec:expcr}
The normalized chemical potential $\widetilde{\mu}$ and the normalized pressure $\widetilde{\Pi}$ is given by $\widetilde{\mu}=\partial f/\partial\eta$ and $\widetilde{\Pi}=\eta\widetilde{\mu}-f$.
Using the minimized free energy expressions of Eqs.\ \ref{eq:fAAA} and \ref{eq:fABC}, which follow from $\partial f/\partial\kappa=0$ and $\partial f/\partial\Delta_\parallel=0$, this gives:
\begin{align}
\widetilde{\mu}&=\frac{f}{\eta}+\frac{2\eta A'}{A}
+\frac{1}{1-\bar{\Delta}_\parallel^2/x},
\label{eq:mcrex}
\\
\begin{split}
\frac{\widetilde{\Pi}}{\eta}&=\frac{2\eta A'}{A}
+\frac{1}{1-\bar{\Delta}_\parallel^2/x},
\label{eq:pcrex}
\end{split}
\end{align}
where $x=\eta_\mathrm{cp,AAA}/\eta$ for the AAA crystal and $x=\eta_\mathrm{ref}/\eta$ for the ABC crystal.

\end{document}